# Basketball scoring in NBA games: an example of complexity.

Y. de Saá Guerra[1], J. M. Martín González[2], S. Sarmiento Montesdeoca[1], D. Rodríguez Ruiz[1], N. Arjonilla-López[1], J. M. García-Manso[1].

[1]Department of Physical Education. University of Las Palmas de Gran Canaria.
[2]Department of Physics. University of Las Palmas de Gran Canaria.

Corresponding author: yvesdesaa@gmail.com (Y. de Saá).

## Abstract

Scoring in a basketball game is a process highly dynamic and non-linear type. The level of NBA teams improve each season. They incorporate to their rosters the best players in the world. These and other mechanisms, make the scoring in the NBA basketball games be something exciting, where, on rare occasions, we really know what will be the result at the end of the game. We analyzed all the games of the 2005-06, 2006-07, 2007-08, 2008-09, 2009-10 NBA regular seasons (6150 games). We have studied the evolution of the scoring and the time intervals between points. These do not behave uniformly, but present more predictable areas. In turn, we have analyzed the scoring in the games regarding the differences in points. Exists different areas of behavior related with the scorea and each zone has a different nature. There are point that we can consider as tipping points. The presence of these critical points suggests that there are phase transitions where the dynamic scoring of the games varies significantly.

## Introduction

All indicate that the degree of competitiveness of a basketball league has a non-linear behavior. This is demonstrated in the work of Yilmaz and Chatterjee (2000) and de Saá et al., (2011). Everything depends upon the equality between teams and the level of uncertainty before and during each game. The evolution of the marker and its final value are those which generate uncertainty for each game and the final standings of a league.

A basketball game is postulated as the clashes between two complex systems (teams) that seek to overcome one over the other in a limited time (Chatterjee and Yilmaz, 1999; Bar-Yam, 2000; Vaz de Melo et al., 2008). In complex systems, processes occur simultaneously at different scales or levels of its components. All are important and reflect the reality of basketball. The intricate behavior of a complex system as a whole depends on its units indirectly, which have strong relationships with each other, often non-linear (Goodwin, 2000, Vicsek, 2002; Amaral & Ottino, 2004; Solé, 2009).

Scoring, a priori, should present a stochastic behavior. The goals are supposed to reflect a completely random dynamic. Should present a random dynamic similar to Brownian motion (random trajectory described by a particle), in the sense that we do not know how big the runs of points will be, or how often. That is, we can not know in advance the dynamic behavior of basketball game scoring.

Far from these assertions, the reality shows us that the score of a basketball game is a direct reflection of the dynamic and non-linear interactions of the teams and its components. However, the evolution of the score seems to have certain patterns or properties that confer identifiable characteristics of each league. Developing a methodology to identify them, allow us to know in detail the internal logic of competition.

Therefore, we consider interesting to study the dynamics of the score of basketball games and, more specifically, its evolution in the games of professional american basketball league (NBA) during 2005-06; 2006-07; 2007-08, 2008-09; 2009-10 seasons.

## Methodology

We studied a total of 5 seasons (1230 games per season, with a total of 6150 games) of the NBA regular season. In every game we analyzed the game transcription published by the NBA in which are described in detail, all incidents that occur play by play (NBA). All the statistics reflect the incidences of game ordered by the time in which they occurred (chronological order): two and three points shots, free throws (made and missed), defensive and offensive rebounds, turnovers and steals; violations (out of bounds, fouls, technical, etc.) and substitutions. From all this information we focus on the analysis of time transcurred between each point that achieved by any team.

If the time-scoring is random, the problem is analogous to a problem of arrivals, and it is modeled by Poisson distribution. In this case the time between baskets follow an exponential distribution, which becomes a straight line on a semilog plot.

Other distributions of interest in complex systems are the *power laws* (PL), which follow many natural phenomena,

often fractal, are also evident in many not natural systems. A lot of elements interact to produce a structure of higher level. These systems evolve far from equilibrium and are often highly dissipative (systems far from equilibrium). The *power laws* are described by mathematical expressions such as:

$$Y = cX^b$$

Where X and Y are two variables, or observable quantities, *c* is a constant (it can also be seen as a normalization constant), and *b* is the scaling exponent. This kind of expression has two properties:

1) The logarithmic transformation becomes a line:

$$log(Y) = log(c) + b\, log(X)$$

2) It is invariant to scale changes.

Phenomena with this type of behavior (*power laws*) are also called scale-free. By scale we mean the spatial and temporal dimension of a phenomenon. The hypothesis of scale that arise in the context of the study of critical phenomena led to two categories of predictions, both have been well verified by a large amount of experimental data on various systems. One of the most important is the scaling law we have mentioned, whose usefulness lies on linking the various critical exponents characterizing the singular behavior of the order parameter and response functions (Amaral & Ottino., 2004).

First we calculated the time between goals and we plotted (Figure 1). Thereafter, to these numerical series obtained from timing differences between goals, was performed a semilog to find out if the behavior was deterministic or not (Figure 2). Then we calculated the difference existing between points in the final score of each game and plotted the data obtained (Figure 3). To these data we applied a log-log to verify whether differences in the score responds to a *power law* (Figure 4).

## Results

The Figure 1 represents the distribution between time and field goal (two-point shots, three points and the first free throw, we must bear in mind that on the second free throw, time still stopped):

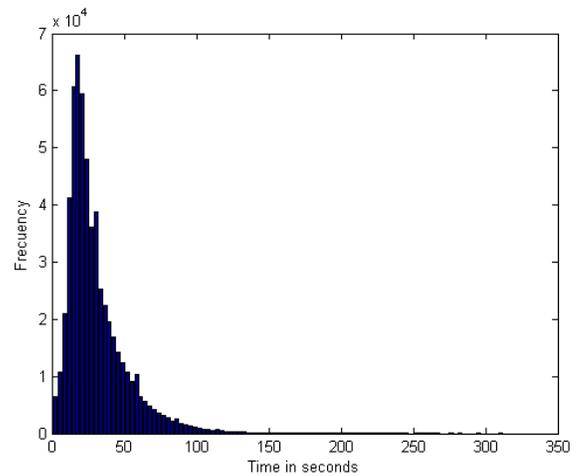

**Figure 1.** Frequencies histogram of the of time between points in the sample analyzed. We can observe that the distribution is not homogeneous. A peak exist around 20´´ time value, and also present a long tail behavior

At the X-axis are represented the time in seconds, in which the goals are produced. At the Y-axis are represented the frecuency of goals. The first thing that stands out is that the data show a not simetric behavior. The distribution has a tail long tail apparently with a maximum value of 310´´, and a frequency peak around the value of 20´´.

The behavior of the tail is best seen by taking logarithms: the Figure 2 shows the time intervals between goals (X-axis) and logarithm of the frequencies (Y-axis). In this case the time between points follow an exponential distribution, which becomes a straight a semilog plot. The upper panel represents the log-log plot of this same data set.

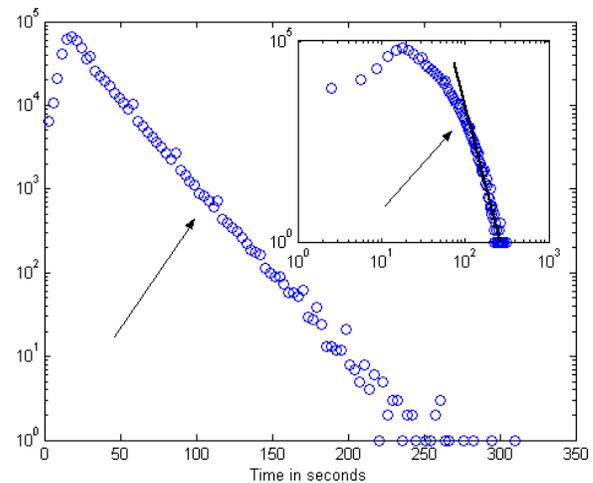

**Figure 2**. Time interval series between points and and logarithm of the frequency. Apparently, there are two different behaviors. Below 24´´ the data are a type of distribution with a maximum (peak) around 20´´. Above 24´´ follows an exponential distribution. To further analyze the behavior of the tail of the distribution (from the 100´´), we also carried out a log-log (upper

panel) to verify how this trend is approaching a *power law* like behavior.

The histogram of the differences in the final score is shown in Figure 3 . The X-axis represents the difference in points between the two teams, and the Y-axis represents its frequency.

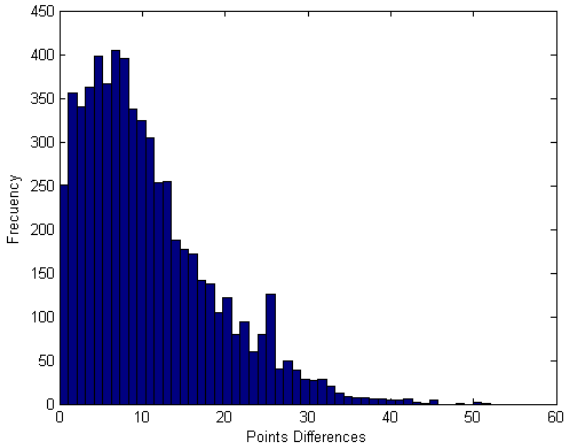

**Figure 3.** Point difference histogram existing in the final score of each game studied. The distribution is approximately uniform for values less than 10-12. From here the distribution shows a possible behavior of long tail.

The Figure 4 represents the log-log plot of difference values in the scoring of the games analyzed. We note as the first data (0 to 10 points approximately) behave almost homogeneously. From the value of 10 points there is an interruption in this trend, indicating a change of behavior on this variable. In addition, exists a second cut on the value of 25-28 points, which again changes its trend. These two lines (last two) appear to indicate the presence of *power laws*.

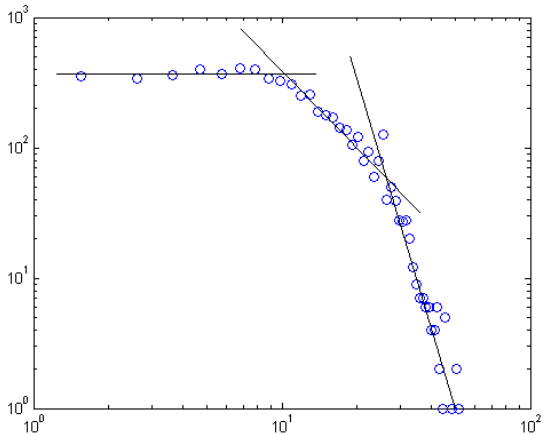

**Figure 4.** Log-log plot of data point difference and frecuency. We can see that the first array present a homogeneous tendency. Around the value of 10 points, an interruption in this trend take place. And a second one at a value around 25-28. This suggests the presence of more than one *power law*.

## Discussion

At the Figure 1 we can see that the distribution has a maximum around 20´´, and a possible long tail from the 100´´ taking a large temporary segment. Approximately up 310´´. That is, there was a situation in which neither teams scored goal after 5´ from the previous goal.

Figure 2 shows a decay in a straight line from 24´´, indicating an underlying Piosson phenomenon, ie, completely random, without memory, for waiting times larger than 24´´.

The Figures 1 and 2 show that in a basketball game, the most likely time between goals are around 20´´. This seems logical considering the 24´´ of possession. Below and above these values the probability drops rapidly, although the effect is much greater for short baskets times (note the asymmetry of the curve).

For higher values of this peak, the probability decreases until attained a certain value, begin to be considered rare phenomena (low probability). For values above 100´´ is possible that it begins to exhibit similar behavior to a *power law*.

This would be an interesting result because if it is a Poisson phenomenon, it colud have a feature called *memorylessness:* (also called evolution without after-effects): the number of goals occurring in any bounded interval of time after time *t* is independent of the number of goals occurring before time *t*. It means that the time in which each point is scored is independent of the previous.

But if the right end of the tail decays less rapidly than does the exponential distribution, mean that long time intervals are followed by long time intervals with a probability slightly higher.

Therefore, the tail end of the *power law* must be studied in more depth possibly through Extremes Values Statistics.

Values below 20´´ has high frequency values (Firgure 1 and Figure 2), which suggests that it can predicted with a small margin of error. In this case, inasmuch as the time intervals are relatively short, fastbreaks could be the source of this trend, because are the goals with the highest success rate.

This may be related with rebounds. Defensive rebounds, because allow to build the fastbreak quickly, and offensive rebounds, because allow to score goals with high success rate, and further elaborate successive attacks. Hence, the strategies of many teams are to make faults to avoid these situations, which create serious disadvantages between a team and the other.

If we consider the absolute value of the score, always grows but do not evolve uniformly. This is a reality which is maintained on all basketball games. Score runs and maximum values achieved by the teams may vary, but always does incrementally.

The absolute points reflect the alternation of the hits of both teams. But what that really sets the dynamics of the game is the point differences between a team and another during the game. Above all at the end of the game.

For that reason, we analyzed the differences on the final score of the whole sample analysis (6150 NBA games). The result indicates that most of the games (65%) ended with a difference of between 1 and 11 points, 33% had a difference of between 11 and 28 points, and only 2% did so with a difference of 28 or more points (Figure 3). To verify whether the data followed a *power law* type distribution, we performed a log-log plot whose result can be seen in Figure 4.

From 1 point to 10 points, the distribution is almost uniform, which correspond with situations of high uncertain. But, if we overcome this score, a behavior appears in the form of *power law* (approximately from 10 points to 28 points). This indicates that the nature of game has changed, and even, if we overcome the barrier of 28 points (a second *power law*), the essence of the game changes radically. We must remember that there is not a fixed criterion to identify non-linear complex systems or self-organized criticality behaviors in sport, but the systematic linking of *power laws* behaviors indicates a possible existence (Savaglio, S. and Carbone, V., 2000; García-Manso and Martín-González, 2008).

The log-log-plot of the distribution of points difference, is broken into several power laws, for certain characteristic values that we consider thresholds or critical points, which means that game dynamic can be characterized by several critical phenomena.

Another issue to consider is the temporal evolution of the difference in points in each game. Some thermodynamic systems are characterized by an order parameter. For example, the density varies smoothly with the temperature. This reverse dependency is due to the cohesive forces prevail over the thermal motion as it decreases. Only at phase transitions (above a critical point), change can be abrupt.

In our case, we can take the time evolution of the score differences as an order parameter and analyze how the critical points, previously defined, work here. Within a particular game if the order parameter remains below 10 points, we can consider that the game is competitive, and the final result unpredictable. If a team exceeds the second threshold (28 points) must be a reaction of the second team, but the game is almost defined. Above the third point, the result is predictable. Logically depend on when the game that exceed these thresholds. That is why the point spread could be considered an order parameter of the system.

This fact makes these critical points work as a percolation threshold. Once passed, the properties of the game change. The order parameter characterises the onset at the phase transition.

The presence of these critical dynamics: the critical slowing down and speeding up (these pertubations), suggests that perhaps we are dealing with a phase transition and critical exponents (Scheffer et al., 2009; Mc Garry et al., 2002). And as consequence the score, and therefore the game, behave as complex non-linear system or as a self-organized system criticaly (SOC).

# Conclusions

We can conclude that as far as intervals between goals are concerned, the time on basketball does not follow a uniform behavior, but there are different behaviors in terms of time ranges. These temporal asymmetries indicate that the basketball score behavior has a non-linear nature. The score is a reflection of the different actions and behaviors resulting from the teams clash. It appears that the teams generate complex non-linear systems that are manifested in the way the score evolves.

Given the high degree of randomness that exists in the most of games, with less than 11 points difference, we could suppose that the majority of the teams are in a state which we can consider as "critical state". A non-equilibrium state where the slightest change will cause a change of "game state", as a difference in the score, or a "phase transition". Therefore, the final result is very difficult to predict (Scheffer et al., 2009; Mc Garry et al., 2002).

The competitive dynamics in the NBA can be considered an example of the *Red Queen* hypothesis proposed by Van Valen (1973): For an evolutionary system, continuing development is needed just in order to maintain its fitness relative to the systems it is co-evolving with. It is a race without end. All competitors need to improve to remain competitive.

As future research lines, it would be interesting to see whether the teams which are complex systems, possess or generate phenomena of learning and memory. And the degree of randomness that exists on the scoring of basketball is due to the chaos that reigns during the basketball game.